\documentclass[10pt,conference,letterpaper]{IEEEtran}
\IEEEoverridecommandlockouts

\usepackage{amsmath,amssymb,amsfonts}
\usepackage{algorithmic}
\usepackage{graphicx}
\usepackage{textcomp}
\usepackage{xcolor}

\def\BibTeX{{\rm B\kern-.05em{\sc i\kern-.025em b}\kern-.08em
    T\kern-.1667em\lower.7ex\hbox{E}\kern-.125emX}}

\usepackage{pbalance}
\usepackage{hyperref}

\makeatletter
\newcommand{\linebreakand}{%
  \end{@IEEEauthorhalign}
  \hfill\mbox{}\par
  \mbox{}\hfill\begin{@IEEEauthorhalign}
}
\makeatother

\begin{document}
\title{Fast and Secure Simultaneous Authentication of Equals for WPA3\\
}

\author{%
\IEEEauthorblockN{João Ferreira}
\IEEEauthorblockA{\textit{IEETA / LASI} \\
\textit{University of Aveiro}\\
Aveiro, Portugal \\
joaop.ferreira@ua.pt}
\and
\IEEEauthorblockN{André Zúquete}
\IEEEauthorblockA{\textit{IEETA / LASI / DETI} \\
\textit{University of Aveiro}\\
Aveiro, Portugal \\
\href{https://orcid.org/0000-0002-9745-4361}{0000-0002-9745-4361}}
\and
\IEEEauthorblockN{Hélder Gomes}
\IEEEauthorblockA{\textit{IEETA / LASI / ESTGA} \\
\textit{University of Aveiro}\\
Aveiro, Portugal \\
\href{https://orcid.org/0000-0001-8443-4196}{0000-0001-8443-4196}}
}

\maketitle

\begin{abstract}
The Simultaneous Authentication of Equals (SAE) protocol, introduced in WPA3, provides robust protection against offline dictionary attacks against a network Pre-Shared Key (PSK) and also protection of session keys from other people knowing  that PSK. However, its high computational cost for the Password Element (PE) derivation makes Access Points (APs) vulnerable to CPU exhaustion Denial-of-Service (DoS) attacks. This paper proposes a resilient architectural modification to SAE. First, we introduce an asymmetric resource cost model that offloads the iterative discovery of cryptographic elements to the client, allowing the AP to maintain a fixed computational load during the handshake. To further mitigate brute-force attempts, we implement a mechanism based on a slow-path key derivation (with variable cost key derivation functions, such as PBKDF2 or Argon2), incorporating a deliberate processing delay on the supplicant. Finally, we introduce a ticket-based mechanism to facilitate efficient re-authentication for known devices, bypassing expensive exchanges while preserving system availability under adversarial conditions. Experimental results demonstrate that this architecture significantly mitigates DoS risks without compromising legitimate network access.
\end{abstract}


\section{Introduction}

Although WPA3-SAE successfully mitigates the long-standing vulnerability of offline dictionary attacks found in WPA2-PSK~\cite{rfc:7664}, it introduces a new attack surface for Denial-of-Service (DoS) attacks~\cite{Vanhoef20}. The ``Hunting-and-Pecking'' algorithm, essential for deriving the Password Element (PE) used in Simultaneous Authentication of Equals (SAE), is inherently CPU-intensive. In a typical deployment, an adversary can easily overwhelm an Access Point (AP) by flooding it with forged authentication commit frames, forcing the AP to perform costly elliptic curve operations and memory allocations for non-existent peers.

This research addresses this imbalance by introducing a mitigation strategy that fundamentally reshapes the SAE handshake's resource requirements. By offloading the iterative search for suitable cryptographic elements to the client (Supplicant), the AP is able to maintain a constant and minimal computational overhead, as it only performs a single-step verification of the values provided by the Supplicant. Simultaneously, the system employs a hardened, time consuming key derivation function, specifically configured to introduce a substantial computational delay, ensuring that authentication requests require a significant investment of computational power from the Supplicant. This combination ensures that while the supplicant must prove its legitimacy through a resource intensive task, the AP remains resilient against CPU exhaustion and flood-based attacks.

\section{Context}

\subsection{The SAE Protocol and Dragonfly Handshake}

\newcommand\SAEctr{\textit{SAE\_ctr}}

The IEEE 802.11-2020 standard introduced SAE to replace the vulnerable Pre-Shared Key (PSK) mechanism~\cite{802.11-2020}. At the core of SAE lies the Dragonfly key exchange~\cite{rfc:7664}, which employs a method known as ``Hunting-and-Pecking'' to derive a common PE in both wireless devices. This PE is a point on an elliptic curve derived from a combination of the PSK, the MAC addresses of both peers and a counter ({\SAEctr}), through a process that ensures that the derivation is tied to the specific session.
However, this is a probabilistic and iterative process; as noted in~\cite{Vanhoef20}, the algorithm running on the AP must perform a minimum of $40$ iterations to ensure resistance against timing attacks. In each cycle, increasing {\SAEctr} ensures a deterministic, yet unpredictable, calculation of PE. This process, which involves quadratic residue tests and hash-to-curve mappings~\cite{rfc:9380}, results in a high computational overhead that is particularly taxing for embedded hardware. On the other hand, the mobile device, which initiates SAE, can perform a faster calculation of PE, since the value to be used is the first valid value found while increasing {\SAEctr}.

\subsection{SAE Messages and Computations}

When a Supplicant wants to connect to an AP, it iterates over {\SAEctr} to find a suitable PE, generates two random values, $p$ (private) and $m$ (mask), and computes $s$ and $E$ from these as follows:
\begin{eqnarray*}
s &=& (p + m) \pmod q \\
E &=& -m \cdot PE
\end{eqnarray*}
where $q$ is the modulo of the elliptic curve group used. The Supplicant then sends $s$ and $E$ in an SAE Commit frame to the AP.
The AP uses the same process to find PE, but naturally uses different random values $p$ and $m$ to compute $s$ and $E$. Then, it also sends an SAE Commit frame to the Supplicant with its $s$ and $E$. By combining the received $s$ and $E$ with their $p$, both parties reach a common secret, which is a fresh PMK (Pairwise Master Key) randomly derived from the PSK. No one, not even other people who know PSK, can guess the value of PMK.

Upon this commit phase, both parties engage in a confirmation phase, where they exchange SAE Confirm frames with a confirmation value, $c$, calculated with a PMK-keyed HMAC over the received and sent $s$ and $E$ values. Since the order of these $4$ parameters varies, the exchanged values $c$ are different. A mutual validation of the $c$ values ensures both that they reached a common PMK.

\subsection{Resource Exhaustion Vulnerabilities}
\label{REV}

While SAE effectively prevents offline dictionary attacks, it introduces a significant vulnerability: CPU exhaustion. The protocol requires the AP to perform intensive computations of elliptic curves for every connection attempt. This creates a perfect scenario for an adversary to flood the AP with forged AES Commit frames.
As discussed in~\cite{Vanhoef20}, an attacker can exploit this by sending a high volume of commit messages, forcing the infrastructure to waste resources in search for a PE for each spoofed MAC address.
Even with the standard anti-clogging defense defined in the 802.11 specification~\cite{802.11-2020}, the AP is not fully protected. As demonstrated in recent security analysis~\cite{Vanhoef20}, anti-clogging only verifies that a Supplicant is ``real'' (MAC reachability), but it cannot stop a motivated attacker from completing the token exchange and forcing the AP to execute the heavy cryptographic handshake. In this scenario, the AP's processor can easily reach 100\% utilization, leading to a DoS where legitimate users are unable to connect.

\subsection{Related Work and Session Resumption Mechanisms}

Standard mechanisms for accelerating wireless reconnections, such as PMK Caching and Opportunistic Key Caching (OKC), were designed to reduce handshake latency by storing pairwise keys from previous sessions~\cite{Lee21}. However, these methods are inherently stateful, requiring the Access Point (AP) to maintain a local database of active keys for every Supplicant. This requirement introduces a different vector for Resource Exhaustion, as an attacker can fill the AP's memory with stale session data. Furthermore, PMK Caching exposes a value, PMKID, which is computed from PSK and several known values (MAC address), which is suitable for exploring off-line dictionary attacks against the PSK~\cite{Steube20}. Finally, exploring PMK Caching prevents a mobile device to using a new, random MAC address on each association to the AP that has the cached PMK.

Another relevant approach is the Fast Initial Link Setup (FILS), introduced in IEEE 802.11ai~\cite{802.11ai-2016}. Although FILS significantly reduces the number of frames exchanged during authentication, it does not fundamentally eliminate the computational burden of the initial SAE key derivation on resource-constrained devices, as it still relies on the heavy ``Hunting-and-Pecking'' process described in~\cite{Vanhoef20}. Our proposed architecture differs from these approaches by implementing a stateless, ticket based resumption. By offloading the session state to the Supplicant within an encrypted envelope, the AP can verify returning devices with minimal CPU and memory overhead, effectively mitigating the vulnerabilities described in Section~\ref{REV}. Furthermore, mobile devices can change their MAC address in consecutive authentications, which is beneficial to reduce their tracking. 

More recently, B.~Lee~\cite{Lee21} proposed a stateless re-association mechanism for WPA3 that uses so-called paired tokens to avoid storing session keys at the AP. Although this approach effectively reduces memory consumption, it introduces a different set of challenges with respect to the management of these paired tokens and the computational cost associated with their verification under high load.

\section{PROPOSED ALTERNATIVE: RESILIENT SAE WITH STATELESS RESUMPTION}

\subsection{Architecture Overview}

The proposed architecture aims to mitigate the CPU exhaustion vulnerabilities identified in Section~\ref{REV} by re-balancing the computational load between the AP and the Supplicant. Our model introduces a bifurcated authentication path: a Resource Intensive Initial Exchange (RIIE) and a Stateless Fast Track Resumption (SFTR).

In RIIE, the Hunting-and-Pecking procedure remains the standard method for the derivation of PE, but its computational burden is strategically redistributed. The Supplicant is required to perform the iterative search for a valid coordinate $x$ on the curve (which yields PE). Once the Supplicant finds PE and sends its Commit frame, the AP only needs to perform a single-step verification of the provided values. Upon a successful RIIE authentication, which ends with the validation of the SAE confirmation messages, the AP issues a cryptographically protected Session Ticket. This ticket encapsulates the session state, allowing the AP to remain stateless. In a subsequent reconnection with SFTR, the Supplicant presents this ticket, enabling the AP to bypass the entire SAE commit-confirm logic. By decrypting the ticket, the AP directly recovers a fresh session key, transforming what would be a complex cryptographic validation into a single, high speed symmetric decryption step.

\subsection{RIIE: Fast PE Derivation and Verification in APs}
\label{riie}

\newcommand\pwdhash{\textit{pwd\_hash}}
\newcommand\HMAC[2]{\text{HMAC}_{#1}\left(#2\right)}

To ensure that the AP remains resilient against flood attacks, with RIIE we redistribute the computational cost of the ``Hunting-and-Pecking'' algorithm.

\begin{figure}[t] 
\includegraphics[width=\columnwidth]{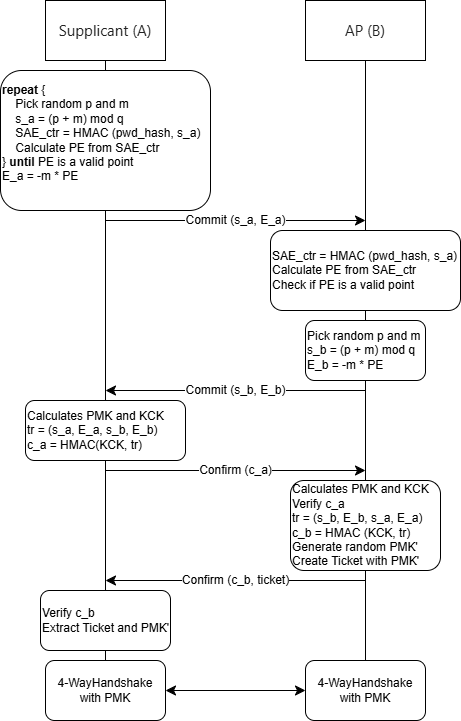}
\caption{Modified SAE protocol that reduces the workload of the AP and distributes a Ticket to the Supplicant }
\label{fig:message_exchange}
\end{figure}

In the standard SAE flow, both the Supplicant and the AP must independently perform an equal iterative search for a common $\SAEctr$ from which they produce their PE. In our model, presented in Figure~\ref{fig:message_exchange}, this search is carried out exclusively by the Supplicant, which then provides the necessary evidence for a single and straightforward calculation and verification of the EP by the AP.

The core of the fast PE derivation by the AP relies on a mapping where {\SAEctr} is generated as:
\begin{eqnarray*}
\text{\pwdhash} &=& \text{H}\left(\text{PSK}\right)\\
\text{\SAEctr} &=& \HMAC{\pwdhash}{s}
\end{eqnarray*}
where $H$ represents some kind of hashing or key-derivation function. The Supplicant uses this {\SAEctr} to compute the PE candidate. If the candidate is not a valid point on the elliptic curve, the Supplicant increments $p$, calculates a new $s$ and repeats the process. Once a valid PE is found, the Supplicant calculates $E$ and sends its SAE Commit frame.

Upon receiving it, the AP's computational load is minimal. Instead of iterating, the AP simply computes the same {\SAEctr} using the provided $s$ and its local copy of (hashed) PSK. It then performs a single-step derivation of PE and verifies its validity, and thus a possible validity of the commitment.

This transformation ensures that the AP's CPU utilization per request remains constant and lower than with the current SAE, regardless of the number of iterations required to find a valid PE (which only occurs on the Supplicant side), effectively mitigating the resource exhaustion attack.

\subsection{Brute Force PSK Discovery Mitigation via Slow Path Derivation}
\label{sec3c}

In the standard ``Hunting-and-Pecking'' algorithm, the iterative nature of the search acts as a natural computational barrier against offline attacks. In our architecture, however, the AP uses a proposed $s$ to derive {\SAEctr} in a single, faster and non-iterative step. While this optimization protects the AP's CPU, it also simplifies the work for an adversary: an eavesdropper who captures the handshake now has a direct path to verify PSK guesses through the $s$ calculation. By testing PSK candidates to generate the observed $s$, an attacker could rapidly verify candidate PSKs without the burden of the original Dragonfly loop.

To counteract this risk, we rely on $H$ to implement a slow path key derivation mechanism. Instead of relying on a single standard, high-speed hashing operation for $H$, {\pwdhash} is generated with a computationally expensive function, such as PBKDF2~\cite{rfc:8018} or Argon2~\cite{rfc:9106}. This process utilizes a number of iterations an a salt that are extracted from the AP's Beacon frames, ensuring that the derivation is uniquely tied to the local network. By making this derivation deliberately slow, we maintain a high computational barrier for adversaries while keeping the AP's verification logic efficient. This ensures that while the AP enjoys a fast, single step verification, the attacker is once again restricted by a significant computational bottleneck, making offline dictionary attacks impractical due to the high cost of testing each individual PSK.

\subsection{SFTR: Stateless Session Tickets and Key Management}
\label{subsec:tickets}

To optimize subsequent connections and ensure that legitimate users only execute the slow path derivation and the iterative ``hunting-and-pecking'' process once for some time, our architecture implements a Stateless Session Ticket mechanism. The security of this mechanism relies on a robust key management flow that ensures confidentiality and authenticity without requiring the AP to maintain local session states.

\subsubsection{Ticket Encryption and Integrity}

Upon a successful initial SAE authentication, the AP generates a PMK to be used by the Supplicant in a future session and encapsulates it in a ticket (PMK' in Figure~\ref{fig:message_exchange}). Then, it encrypts the future PMK with the current one and sends the result, together with the ticket, to the Supplicant. In a future authentication attempt, the Supplicant can redeem the ticket and use the PMK on it, instead of running SAE again. This is similar to the distribution of session keys and tickets in Kerberos~\cite{rfc:4120} and TLS v1.3~\cite{rfc:8446}.

The ticket is protected through a secure encapsulation mechanism that ensures both the confidentiality and integrity of its payload. The AP uses a random, locally stored Secret Ticket Encapsulation Key (STEK) to encrypt the tickets and validate their contents. Different parts of STEK are used for encryption and integrity control.
Since this key is known exclusively to the AP, the Supplicant cannot tamper with or disclose tickets' contents.

The ticket payload includes the following fields, encrypted with STEK:
\begin{itemize}
\item A sequential ticket identifier (TID);
\item A future, random PMK.
\end{itemize}
This payload is protected against tampering and forgery by a STEK-computed HMAC of its encrypted contents, following an Encrypt-Then-MAC paradigm. This benefits fast AP validation, which is fundamental for mitigating DoS attacks.

These tickets allow the AP to verify their authenticity and recover a fresh session PMK without maintaining any records of issued tickets. To further enhance security, namely to reduce DoS attacks using old tickets, tickets should be used at most once. To do so, APs can implement a strategy similar to the one recommended for IPSec~\cite{rfc:4301}: an anti-replay policy to validate TIDs using a sliding acceptance window. Furthermore, APs can implement STEK refresh policies to revocate all unused tickets. This revocation is relevant upon a modification of the AP's PSK, since such action could have been used to evict some users from the network, and their possession of tickets would ruin the intended outcome.

Following a successful ticket-based authentication, the AP can send another ticket to the Supplicant. The principle is the same, but this time the AP will use the ticket's PMK instead of a PMK derived with SAE to protect the future PMK that it sends to the Supplicant.

\subsubsection{Key Recovery and Derivation}

When a Supplicant attempts to reconnect with a ticket, it uses a modified SAE Commit frame, with a null $s$ and the ticket instead of $E$. The AP validates the ticket and, upon success, responds with an SAE Commit frame also with $s = 0$ and a random $E$. Otherwise, it responds with a $s = 1$, and terminates the SAE handshake with the Supplicant. Note that in SAE, the value of $s$ must be higher than $1$, thus there is no confusion with the normal SAE protocol.

If the ticket is accepted, then both the Supplicant and the AP exchange the SAE Confirm messages calculated with the ticket's PMK (instead of the KCK used in the normal SAE and in our modified SAE) and the key agreement protocol proceeds to the ordinary 4-Way Handshake protocol, also with the PMK, which finalizes all Supplicant-AP mutual authentication and key distribution protocol runs.

\section{Implementation}
\label{sec:implementation}

The proposed architecture was implemented as a complete session management system using the \texttt{hostapd v2.10} code base\footnote{https://git.w1.fi/cgit/}. 
The modified source code is publicly available at the following repository: https://github.com/joaopferreira01/fast-sae-wpa3.


\subsection{Access Point Signaling via Information Elements}

\begin{figure}[ht] 
\includegraphics[width=\columnwidth]{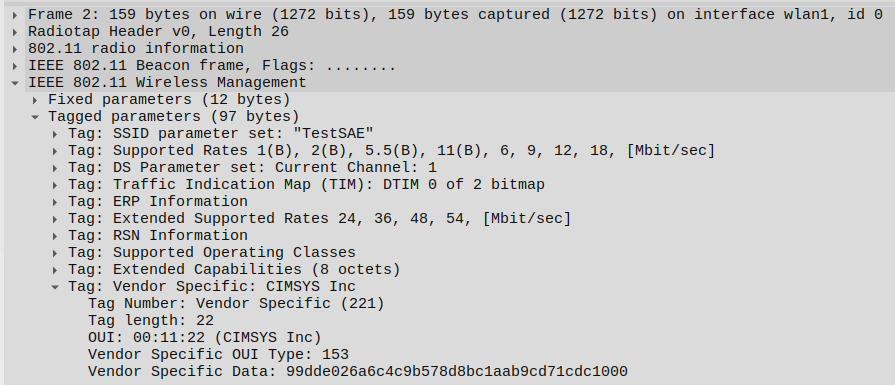}
\caption{Beacon frame captured with WireShark. The last IE contains the salt (\texttt{0x99b7}$\cdots$\texttt{51d1}) and number of iterations (\texttt{0x1000}, or $4096$). These values are used by the Supplicant to calculate {\pwdhash} from the network passphrase with PBKDF2.}
\label{fig:beacon_frame}
\end{figure}

To support the bifurcated authentication path (RIIE and SFTR), the AP must signal its capabilities to the supplicants. This was implemented by modifying the \texttt{hostapd} beacon generation logic to include a custom Information Element (IE). We used a vendor-specif IE with a data value where the first 16 bytes have the salt and the last 2 bytes the number of iterations (see Figure~\ref{fig:beacon_frame}). This IE is broadcast in the Beacon and Probe Reply frames, allowing Supplicants to identify that the network implements the modified SAE protocol and the ticket-based resumption. This IE indicates the iterations and the salt to be used with PBKDF2 to derive {\pwdhash} from the passphrase that produces PSK.

\subsection{State Machine and Message Modifications}

The \texttt{hostapd} authentication state machine was modified to handle the bifurcated path described in Sections~\ref{riie} and~\ref{subsec:tickets}. We introduced a "fast-path" handler that intercepts SAE Commit frames. If the frame contains a ticket (indicated by $s = 0$, as described in Section~\ref{subsec:tickets}), the system bypasses the calculation logic of {\SAEctr} from $s$ and proceeds directly to ticket validation and PMK recovery.

Tickets are distributed by the AP to Supplicants as an extra field of its SAE Confirm frame.
In WireShark, we could see them as extra bytes after the regular value that exists in that frame.


\section{Security Analysis}
\label{sec:security}

The security of the proposed architecture relies on the robustness of the session ticket mechanism and the efficiency of the verification path at the AP.

\subsection{Resilience Against DoS attacks}

The primary security objective is to protect the AP's CPU from exhaustion. Using a cached {\pwdhash} for PE derivation and a lightweight HMAC for ticket verification, the AP avoids performing the heavy derivation process of {\SAEctr} from $s$ or the recommended iterative ``Hunting-and-Pecking'' procedure.

\subsection{Ticket Integrity, Confidentiality and Replay Protection}

The security of the stateless session ticket is guaranteed by STEK, which is known only to the AP. By encrypting the ticket payload with STEK, we ensure that the future PMK remains opaque to eavesdroppers during session resumptions. Any attempt to tamper with the ticket, such as modifying its TID or PMK, will result in an integrity control failure, causing the AP to reject the request.

Regarding replay protection, the inclusion of a sequential TID and the recommended anti-replay sliding window (as mentioned in Section~\ref{subsec:tickets}) prevents anyone, both legit Supplicants and eavesdroppers, from successfully reusing tickets.

\subsection{Resistance to Offline Brute-Force Attacks}

Although the AP uses a fast verification path, the initial derivation of {\pwdhash} remains protected by a computationally expensive function (H). As described in Section~\ref{sec3c}, this ensures that an attacker who captures the $s$ value in a Supplicant's SAE Commit frame still faces a high computational barrier to implement a brute-force discovery of PSK.

\subsection{Forward Secrecy }

SAE forward secrecy guaranties are due to the use of random values $p$ and $m$ in the calculation of $s$ and $E$. Since our SAE version maintains this -- we only derive PE from the Supplicant's $s$ value --, we keep the forward secrecy guaranties.

However, this is not true when the Supplicant uses tickets, since a future PMK is distributed to a Supplicant encrypted with the current one. Furthermore, tickets are encrypted with STEK, which is a key that can stay constant for a long time, which also destroys forward secrecy guaranties.
This loss of forward secrecy when tickets are used is a price to pay for faster reauthentications when the Supplicant wants to randomize its MAC address for increasing its privacy.

Nevertheless, if forward secrecy is fundamental, a genuine Supplicant can derive {\pwdhash} from PSK only once and speed-up their discovery of {\SAEctr}. Furthermore, a Supplicant not willing to randomize its MAC address can reauthenticate almost as fast as with tickets by reusing its previous $s$ value, while changing $p$ and $m$ that are used in its computation (it is enough to increment one and decrease the other with the same random value). This means that $E$ needs to be recalculated, but not $s$, which is the element that needs to be found by trial and error.

\subsection{Protection Against Parameter Downgrade Attacks}
Since the AP imposes the parameters for the password derivation (such as the salt and the number of iterations), a rogue AP could potentially attempt a downgrade attack by sending a low iteration count. This would reduce the computational cost of the PE derivation, making it easier for the attacker to perform an offline dictionary attack on the captured $s$ value. To mitigate this threat, Supplicants must implement a local security policy that enforces a minimum iteration threshold (e.g. $10^8$). By rejecting any parameters below this threshold, Supplicants ensure that they cannot be forced to generate weak cryptographic values, thus maintaining the integrity of the high computational barrier.

\section{Performance Evaluation}
\label{sec:evaluation}

This chapter presents experimental results obtained from our proposed architecture. We focus on comparing the authentication latency and computational overhead of our model with the standard WPA3-SAE implementation.

\subsection{Experimental Setup}
\label{subsec:setup}

To evaluate our protocols, a controlled testbed was established using a virtualized network environment. This setup isolates the computational cost of cryptographic modifications from external wireless interference and network jitter.

\begin{itemize}
    \item {Hardware Infrastructure:} The host machine is equipped with an {Intel Core i9-13900HX} processor (2.20 GHz, up to 5.40 GHz) and 32 GB of DDR5 RAM. Dedicated resources were allocated to two independent Virtual Machines (VMs) to simulate the Access Point (AP) and the Supplicant.
    \item {Software Environment:} The prototype was developed by modifying the official source code of {hostapd v2.10} (for the AP) and {wpa\_supplicant v2.10} (for the Supplicant); these implementations were linked against the {OpenSSL} library to handle Elliptic Curve Cryptography (ECC) and the PBKDF2 algorithm.
    \item {Implementation Details:} For the initial "Slow Path" authentication, the PBKDF2 derivation was configured with a baseline of {10 iterations}. While this value serves for functional validation in the current prototype, it was designed to be scaled to higher iteration counts to meet specific security requirements.
    \item {Network Configuration:} The VMs were interconnected via a virtual bridge, prioritizing the measurement of processing latency over transmission time.
\end{itemize}

\subsection{Authentication Latency Analysis}

In this section, we analyze the impact of our modifications on the time required to establish a secure connection. Latency was measured from the initiation of the SAE exchange until the derivation of the PMK.

\begin{table} [b]
\centering
\caption{AP Computational Performance: Standard vs. Proposed Model ($N=100$)}
\label{tab:performance_results}
\begin{tabular}{lrr}
\hline
\textbf{Authentication Scenario} & \textbf{Mean ($\mu$s)} & \textbf{Std. Dev. ($\sigma$)} \\ \hline
Standard SAE (Hunting-and-Pecking)     & 2284 & 521.8 \\
Proposed Model (Initial SAE auth.)    & 72.5  & 11.0  \\
Proposed Model (fast-path w/ ticket) & 21   & 4.9   \\ \hline
\end{tabular}
\end{table}

\subsubsection{Initial Authentication Performance}

As shown in Table~\ref{tab:performance_results}, the standard SAE ``Hunting-and-Pecking'' mechanism presents a significant computational cost, with a mean latency of 2284~$\mu$s and high temporal instability ($\sigma = 521.8$). In contrast, the proposed SAE authentication strategy reduces the mean processing time to 72.5~$\mu$s, representing a reduction of approximately 96.8\% in computing effort. Furthermore, the significantly lower temporal variance ($\sigma = 11.0$) indicates a more consistent and predictable execution time compared to the iterative nature of the standard SAE process.

\subsubsection{Re-authentication: The Efficiency of Stateless Tickets}

The most significant performance gain was observed during re-authentications with tickets. By using the stateless ticket, the Supplicant and the AP were able to bypass the PE derivation entirely.
As presented in Table~\ref{tab:performance_results}, we observed a reduction of approximately 99\% in authentication latency. Since the AP only needs to perform a single symmetric decryption, the overhead is negligible, proving the scalability of our architecture for mobile Supplicants.

\subsection{Final Remarks}

The experimental evaluation demonstrates that the proposed architecture successfully addresses the primary limitations of the standard WPA3-SAE handshake. By shifting from a probabilistic iterative search to a deterministic key derivation, we achieved a more stable and predictable resource consumption profile in the AP. 

The results indicated that while the slow-path mechanism for derive {\pwdhash} maintains the necessary security barriers against offline dictionary attacks, the introduction of stateless session tickets provides a high-performance alternative for subsequent authentications. This optimization effectively eliminates the heavy computational burden of elliptic curve cryptography during re-authentications, reducing latency to a fraction of the original requirement. Ultimately, the data confirm that it is possible to maintain the high security standards of WPA3 while significantly improving the efficiency and scalability of the authentication process for resource-constrained network infrastructures.

\section{Conclusions and Future Work}
\label{sec:conclusion}

This paper presents an alternative architecture for the WPA3-SAE authentication protocol, specifically designed to mitigate the computational imbalance inherent in the ``Hunting-and-Pecking'' algorithm. By redesigning the Password Element (PE) finding process, we successfully shifted the heavy computational workload away from the APs, ensuring that infrastructure resources are preserved without compromising the security of the network. The implementation of a deterministic slow-path key derivation from the network passphrase to broadcast $s$ values ensures that the protocol remains resilient against offline dictionary attacks, maintaining the core security promises of the Dragonfly exchange.

The main contributions of this research are four-fold:
\begin{enumerate}
    \item {Resource-efficient SAE:} We demonstrated that replacing probabilistic iterative loops with direct key derivation significantly stabilizes CPU utilization on the AP, making WPA3 more viable for low-power IoT gateways.
    \item {High-performance re-authentication:} Through the introduction of stateless session tickets, we achieved a drastic reduction in reconnection latency, enabling very fast transitions for mobile devices.
    \item {Scalable security architecture:} Our model provides a stateless approach to key management, allowing network infrastructures to handle a higher density of concurrent authentication attempts without the risk of memory exhaustion or DoS attacks.
    \item {Adaptable security requirements:} by adding the arbitrary parameters for the PBKDF2 key derivation to a Beacon IE, we allow network administrators to impose an arbitrary burden on passphrase-guessing attacks, enabling them to cope with the evolution of technology without changing standards or implementations. 
\end{enumerate} 

Although the results are promising, several avenues for future research remain to be explored. Firstly, regarding hardware integration, future efforts should focus on deploying this modified \texttt{hostapd} implementation on commercial embedded routers to evaluate performance under real-world AP's hardware constraints. Furthermore, investigating methods to maintain backward compatibility with standard WPA3 Supplicants while offering the optimized path to supported devices would be a significant step towards industrial adoption. Finally, the design of new tickets that could provide fast PMK installation while ensuring forward secrecy.


\section*{Acknowledgment}

This work was supported by the Foundation for Science and Technology (FCT) through contract \href{https://doi.org/10.54499/UID/00127/2025}{doi.org/10.54499/UID/00127/2025}.

\bibliographystyle{IEEEtran}
\bibliography{bib}


\end{document}